\documentclass[twoside, 12pt]{article}

\usepackage{lipsum} 

\usepackage[sc]{mathpazo} 
\usepackage[T1]{fontenc} 
\usepackage{microtype} 

\usepackage[hmarginratio=1:1,top=32mm,columnsep=20pt]{geometry} 
\usepackage[hang, small,labelfont=bf,up,textfont=it,up]{caption} 
\usepackage{booktabs} 
\usepackage{float} 
\usepackage{hyperref} 

\usepackage{lettrine} 
\usepackage{paralist} 

\usepackage{abstract} 

\usepackage{titlesec} 
\renewcommand\thesection{\Roman{section}} 
\renewcommand\thesubsection{\Roman{subsection}} 
\titleformat{\section}[block]{\large\scshape\centering}{\thesection.}{1em}{} 
\titleformat{\subsection}[block]{\large}{\thesubsection.}{1em}{} 

\usepackage{bm}
\usepackage{graphics}
\usepackage{graphicx}
\usepackage{color}

\usepackage{fancyhdr} 
\title{\vspace{-15mm}\fontsize{16pt}{10pt}\selectfont\textbf{Nuclear Pairing from Two-body Microscopic Forces:
Analysis of the Cooper Pair Wavefunctions}} 

\author{
\large
\textsc{Paolo Finelli}\\[1mm] 
\normalsize{Department of Physics and Astronomy, University of Bologna (Italy)} \\
\normalsize{and INFN, Bologna section (Italy)}
~\vspace{5mm}\\
\large
\textsc{Stefano Maurizio}\\[1mm] 
\normalsize{Albert Einstein Center for Fundamental Physics,}\\
\normalsize{Institute for Theoretical Physics, Bern
(Switzerland)}
~\vspace{5mm}\\
\large
\textsc{Jeremy W. Holt}\\[1mm] 
\normalsize{Department of Physics, University of Washington, Seattle (USA)} 
}
\date{}

\begin{document}

\maketitle 

\thispagestyle{fancy} 


\begin{abstract}

\noindent 
 In a recent paper \cite{ourpaper} we studied the behavior of the pairing gaps $\Delta_F$ as a function of 
the Fermi momentum $k_F$ for neutron and nuclear matter in all relevant angular momentum 
channels where superfluidity is believed to naturally emerge. The calculations
employed realistic chiral nucleon-nucleon potentials \cite{Machleidt:2011zz, Epelbaum:2008ga}
with the inclusion of three-body forces and self-energy effects. In this contribution, after a detailed description
of the numerical method we employed in the solution of the BCS equations, we will show a preliminary analysis of the Cooper pair wavefunctions.
\end{abstract}



\section{Introduction}
\label{intro}
The goal of this article is to solve the BCS equations starting from modern
nucleon-nucleon potentials (at N3LO in the chiral expansion \cite{Machleidt:2011zz}) and to perform preliminary calculations of the Cooper pair wavefunctions as a first step towards an analysis of the occurrence of BCS-BEC crossover in nuclear systems \cite{futurepaper}. 

\section{Khodel's method}
\label{sec-1}
In this section we explain the method suggested in Ref. \cite{Khodel S} to solve the BCS equations by 
partial-wave decomposition that has been proven to be stable even for small values of the gap and 
to require only the initial assumption of a scale factor $\delta$ (results, of course, will be 
$\delta$-independent).
The BCS equation reads in terms of the NN potential $ V(\textbf{k}, \textbf{k}')  =
\langle{ \textbf{k} \left| V\right|  \textbf{k}'  }\rangle$ as follows
\begin{equation}
\Delta \left(\textbf{k}\right) = -\sum_{\textbf{k}'} \langle{ \textbf{k} \left| V\right|  \textbf{k}'  }
\rangle \frac{ \Delta \left(\textbf{k}'\right)}{ 2 E\left( \textbf{k}' \right)} \; ,
\label{gapeq}
\end{equation}
with $ E(\textbf{k})^2=\xi(\textbf{k})^2+\left|\Delta(\textbf{k})\right|^2$ and where 
$\xi(\textbf{k})=\varepsilon(\textbf{k})-\mu$, $\varepsilon(\textbf{k})$ denotes the single-particle 
energy and $\mu$ is the chemical potential. We can decompose both the interaction and the gap function
\begin{eqnarray}
\langle{ \textbf{k} \left| V\right|  \textbf{k}'  }\rangle & = & 4 \pi \sum_l (2 l +1) P_l(\hat{\textbf{k}}\cdot 
\hat{\textbf{k}}') V_l(k,k') \\
\Delta(\textbf{k}) & = & \sum_{l m} \sqrt{\frac{4 \pi}{2 l +1}}Y_{l m}(\hat{\textbf{k}})\Delta_{l m}(k) \, ,
\end{eqnarray}
where $Y_{l m}(\hat{\textbf{k}})$ denotes the spherical harmonics, $ l$ and $m$ are the quantum numbers associated with the orbital angular momentum and its projection along the $z$ axis and $ P_l(\hat{\textbf{k}}\cdot \hat{\textbf{k}}')$ 
refers to the Legendre polynomials. 
After performing an angle-average approximation we have the following equation for any value of $l$
\begin{eqnarray}
\label{eq:AAgap}
\Delta^j_l(k)=\sum_{l'}\frac{ (-1)^\Lambda}{\pi}\int{dk'~V^j_{l l'}(k,k')}\frac{\Delta^j_{l'}(k')}{E(k')}{k'}^2 \, ,
\end{eqnarray}
where $\Lambda =1+ (l-l')/2$, $j$ refers to the total angular momentum ($\bm {J} = \bm{l} + \bm{S}$) 
quantum number including spin $\bm{S}$ and now 
$E(k)^2 =\xi(k)^2+ \sum_{j l}\Delta^j_l(k)^2$. Gaps with different $l$ and $j$ are coupled due to 
the energy denominator but we assume that different components of 
the interaction mainly act on non-overlapping intervals in density. 

\begin{figure}[ht]
\centering
\includegraphics[width=1.0\textwidth,natwidth=610,natheight=642]{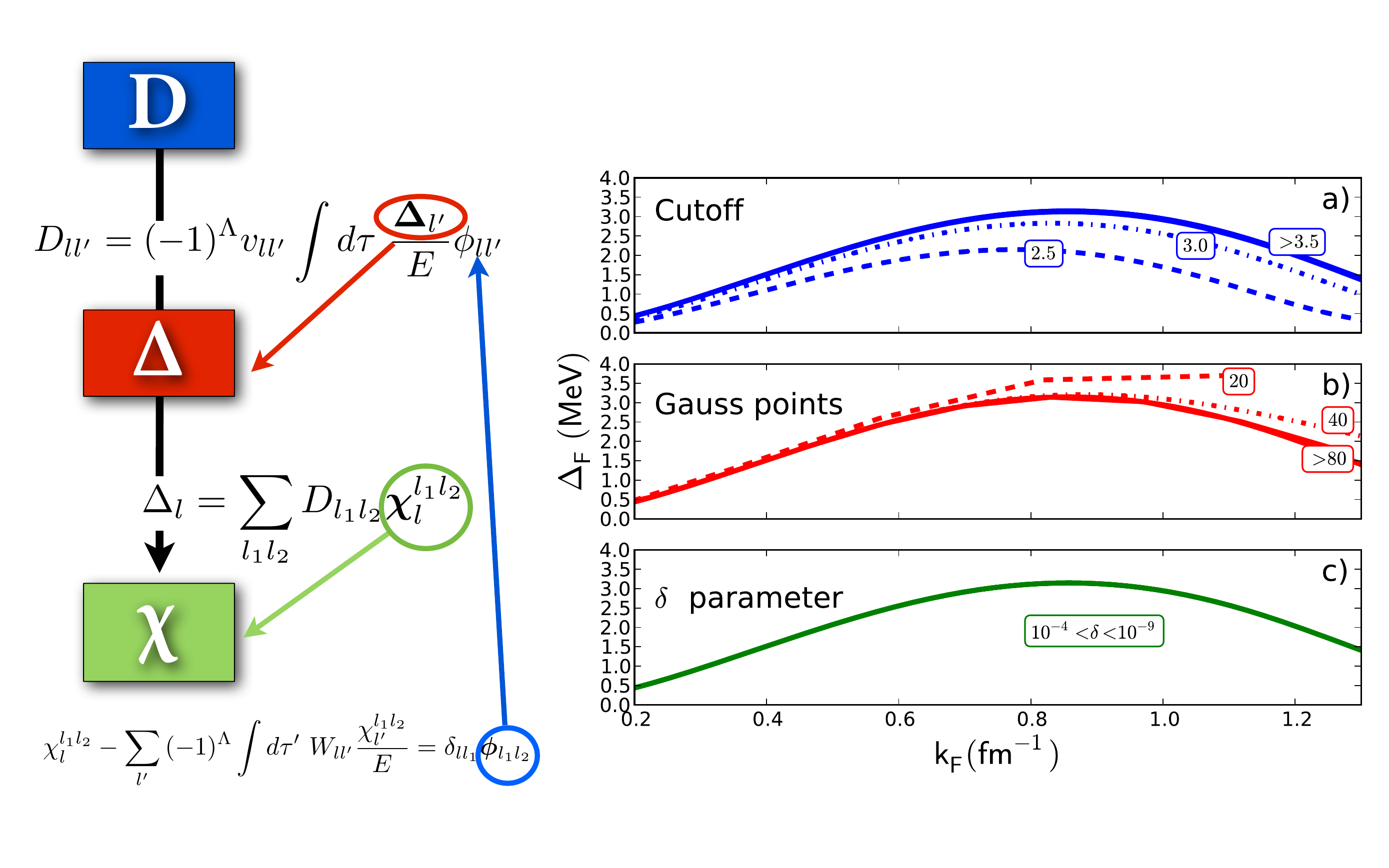}
\caption{{\bf Left}: Self-consistent procedure (Eqs. \ref{eq:coeff}--\ref{eq:functions}) for the solution of the gap equation according to Khodel's prescription \cite{Khodel S}.
{\bf Right}: Numerical analysis of Khodel's procedure for the singlet channel in neutron matter: {\bf a)} cutoff, {\bf b)} Gaussian integration points and {\bf c)} $\delta$ dependence. This method is a very stable procedure if satisfactory values 
of $n_{gauss}$ and $\Lambda_k$ are employed.}\label{fig1}
\end{figure}

We define an auxiliary potential $W$ according to
\begin{eqnarray}
W_{l l'}(k,k') = V_{ll'}(k,k') - v_{ll'}\phi_{ll'}(k)\phi_{ll'}(k') \; ,
\end{eqnarray}
where $ \phi_{ll'}(k)=V_{ll'}(k,k_F)/V_{ll'}(k_F,k_F) $ and $v_{ll'}=V_{ll'}(k_F,k_F) $                       
so that $W_{ll'}(k,k')$ vanishes on the Fermi surface.
The coupled gap equations can be rewritten as
\begin{eqnarray}
\Delta_l(k)-\sum_{l'}{ (-1)^\Lambda \int{d\tau'~  W_{ll'}(k,k')\frac{\Delta_{l'}(k')}{E(k')}=
\sum_{l'}{D_{ll'} \phi_{ll'}(k)}   } } \; ,
\end{eqnarray}
where $d\tau=k^2 dk/\pi$ and the coefficients $D_{ll'}$ satisfy
\begin{eqnarray}
\label{eq:coeff}
D_{ll'}=(-1)^\Lambda v_{ll'} \int{d\tau ~\phi_{ll'}(k) \frac{\Delta_{l'}(k)}{E(k)}} \; .
\end{eqnarray}
The gap is defined as follows
\begin{eqnarray}
\label{eq:buildgap}
\Delta_l(k)=\sum_{l_1l_2} D_{l_1l_2}\chi^{l_1l_2}_l(k) \; ,
\end{eqnarray}
where
\begin{eqnarray}
\label{eq:functions}
\chi^{l_1l_2}_l(k)-\sum_{l'}{(-1)^{\Lambda} \int   {d\tau'~ W_{ll' }(k,k') \frac{ \chi^{l_1l_2}_{l'}(k')}{E(k')} } }
=\delta_{ll_1}  \phi_{l_1l_2}(k) \; .
\end{eqnarray}
The property that $W_{ll'}(k,k') $ vanishes on the Fermi surface ensures a very weak 
dependence of $ \chi^{l_1l_2}_l(k)$ on the exact value of the gap so that, in first approximation, 
it is possible to rewrite the previous equation (\ref{eq:functions}) as
\begin{eqnarray}
\label{eq:functionsFO}
\chi^{l_1l_2}_l(k)-\sum_{l'}{(-1)^{\Lambda} \int   {d\tau'~ W_{ll' }(k,k') \frac{ \chi^{l_1l_2}_{l'}(k')}{\sqrt{{\xi^2(k')}+\delta^2}} } }
=\delta_{ll_1}  \phi_{l_1l_2}(k) \; .
\end{eqnarray}
We use this equation to evaluate $\chi^{l_1l_2}_l(k)$ initially by matrix inversion, then we use 
this function to self-consistently evaluate $D_{ll'}$. Finally, we solve the system given by 
Eqs.\ (\ref{eq:coeff})--(\ref{eq:functions}) in a self-consistent procedure as shown in Fig. \ref{fig1} (left panel).
We always assume $\mu = \varepsilon_F$ and adopt the relativistic version of the single-particle 
energy $\varepsilon \left( k \right) = \sqrt{k^2 + M_N^2 }$, where $M_N$ is the nucleon mass.
 
\paragraph{Numerical analysis}
\label{sec-12}
~\\
As a benchmark, we will consider the $^1$S$_0$ pairing gap in neutron matter, but the same conclusions can be drawn for all interaction channels. Concerning the singlet channel, at the two-body level, we find a good agreement with the gap computed from well known realistic potentials like the CD-Bonn or Nijmegen interactions \cite{Jensen}, except for
larger densities where the N3LO gap exhibits a higher value (phase shifts from the chiral N3LO potential
exhibit more attraction than the CD-Bonn potential for high momenta \cite{Vlowbcs}). 

We tested Khodel's method \cite{Khodel S} against the variation of the following three parameters: 
$n_{gauss}$ (number of Gauss integration points), $\Lambda_k$ (cutoff for integrals in momentum space, see Eq. (\ref{eq:AAgap})) and $\delta$ (the scale factor). 
In Fig. \ref{fig1} (right side) we summarise our results. In the upper panel ({\bf a}) we calculated $\Delta_F$ 
for different values of the momentum cutoff (using $n_{gauss} = 200$ and $\delta =1\times 10^{-10}$ MeV) where in the second panel ({\bf b}) we varied $n_{gauss}$ (keeping $\Lambda_k = 4.5$ fm$^{-1}$ and $\delta =1\times 10^{-10}$ MeV)
and in the lower panel ({\bf c}) we changed $\delta$ (with $n_{gauss} = 200$ and $\Lambda_k = 4.5$ fm$^{-1}$)
by orders of magnitude.

Our conclusion is that the method proposed by Khodel \cite{Khodel S} is a very stable procedure to study nuclear superfluidity if a reasonable number of Gaussian points ($\ge 100$) and a realistic momentum cutoff  ($\ge 4$ fm$^{-1}$) are employed.

\section{Cooper pair wavefunctions}
\label{sec3}

Recently it has been shown that, at low density, nuclear matter can be subjected to a phase transition, belonging to the BCS-BEC crossover phenomenon \cite{BCSBEC},
caused by the collapse of the Cooper pairs in deuteron-like particles. 
In this phase, the nucleons forming the pairs are strongly correlated; this correlation gives an extra binding energy to the nuclear equation of state and could lead to identify  new properties at low density. 
Because BCS equations are still valid in the BEC regime
\cite{Pieri}, it is useful to study the evolution of pairs of correlated nucleons as a function of density and spatial coordinate, starting from the solution of Eqs. (\ref{eq:coeff}--\ref{eq:functions}).

The Cooper pair wavefunction is defined as follows
\begin{equation}
\Psi_{pair} \equiv C^{'} \langle \Phi_0 | \psi^\dagger ({\bm r}, \uparrow) \psi^\dagger ({\bm r}^{'}, \downarrow) 
| \Phi_0 \rangle =  \frac{C}{(2\pi)^3} \int {\rm d}{\bm k}~ \frac{\Delta (k)}{2E(k)} e^{i {\bm k} \cdot ({\bm r} - {\bm r}^{'})} \; ,
\end{equation}
where $|\Phi_0\rangle$ is the BCS ground state, $\psi^\dagger$ is the particle creation operator and $C, C^{'}$ are normalization factors that can be fixed by imposing normalization conditions.
In the following discussion we will indicate as $\rho(r)$ the probability density to find the nucleons forming the pair at a distance $r$, $P(r)= \int_0^r dr' \rho\left(r'\right)$ where $\rho(r) =|\Psi(r)|^2 r^2 $
in the singlet channel. Observing that the pairing gap in the  $S$ channel is larger then the one in the $D$ channel we assumed, in the SD channel,
$\left|\Psi(r)\right|^2 \approx \left|\Psi_{l=0}(r)\right|^2$ and we have approximated $\Delta (k) $ with $\Delta_{l=0}  (k)$. 
Analogously, in the PF channel, $\Delta (k) \simeq \Delta_{l=1}  (k)$.

As suggested by Matsuo \cite{Matsuo} the spatial behaviour of the Cooper pair varies strongly with the Fermi momentum $k_F$ (or density $\rho_0 = g/(2\pi)^3 k_F^3$, where $g=1$ for neutron matter and $g=2$ in symmetric nuclear matter), in particular the $\rho$-profile shows strong variations. The weak-coupling limit is known to lead to an exponential falloff convoluted with an oscillation suggesting a large correlation length. On the other hand, a pronounced peak with small oscillations could be interpreted as a sign
of a transition to a different regime, i.e. the so called BCS-BEC crossover. Using phenomenological pairing interactions,
Matsuo \cite{Matsuo} suggested that in the singlet channel, over a wide range of densities $(\rho/\rho_0 \simeq 10^{-4} - 0.5)$, the spatial dineutron correlation is strong and a possible crossover region could be found in the density region 
$\rho/\rho_0 \simeq 10^{-4} - 10^{-1}$. 

In the following paragraphs we summarise our results obtained with microscopic two-body NN forces.

\begin{figure}[t]
\centering
\includegraphics[width=0.95\textwidth,natwidth=610,natheight=642]{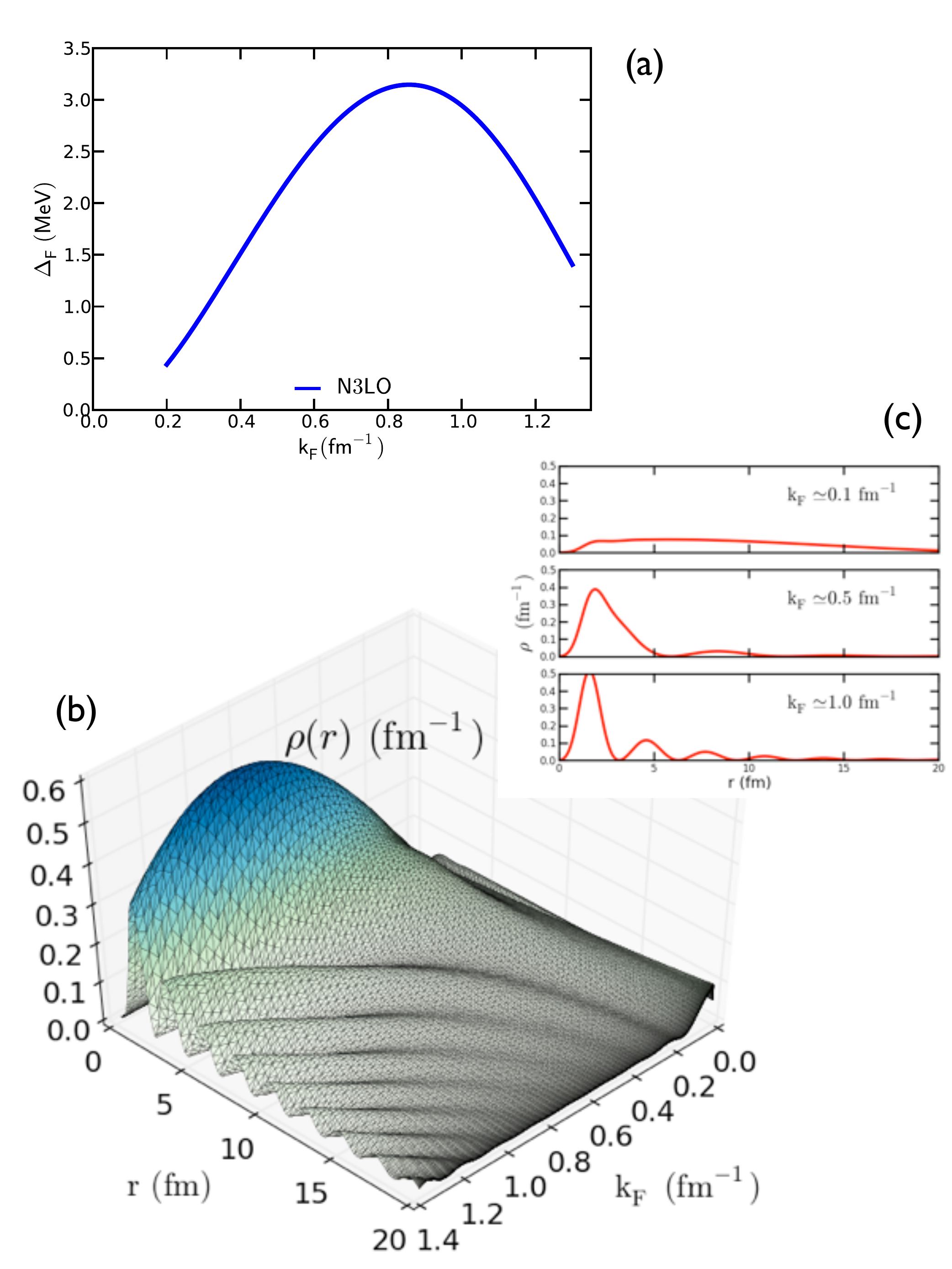}
\caption{{\bf (a)} Pairing gap $\Delta_F$ as function of the Fermi momentum $k_F$ with N3LO interaction \cite{ourpaper} for the singlet channel in neutron matter. {\bf (b)} Probability density $\rho$ as function of the coordinate $r$ and Fermi momentum $k_F$. {\bf (c)} Snapshots of $\rho(r)$ for some selected values of $k_F$.}\label{fig2}
\end{figure}

\paragraph{Singlet case ($^1$S$_0$ in neutron matter)}
~\\
In Fig. \ref{fig2} we show in the upper panel {\bf (a)} the pairing gap $\Delta_F$ as a function of the Fermi momentum $k_F$  \cite{ourpaper} for the singlet channel in neutron matter, where in the lower panel {\bf (b)} the probability density $\rho$ is plotted as a function of the coordinate $r$ and Fermi momentum $k_F$.
For the sake of clarity we also show $\rho(r)$ for few selected values of $k_F$ in the graph labelled by {\bf (c)}. Results of Ref. \cite{Matsuo} are substantially confirmed: at densities below $k_F \simeq 0.8 \mbox{ fm}^{-1}$ the system belongs to a crossover region, while at higher densities is in the pure BCS region.

\begin{figure}[t]
\centering
\includegraphics[width=1.0\textwidth,natwidth=610,natheight=642]{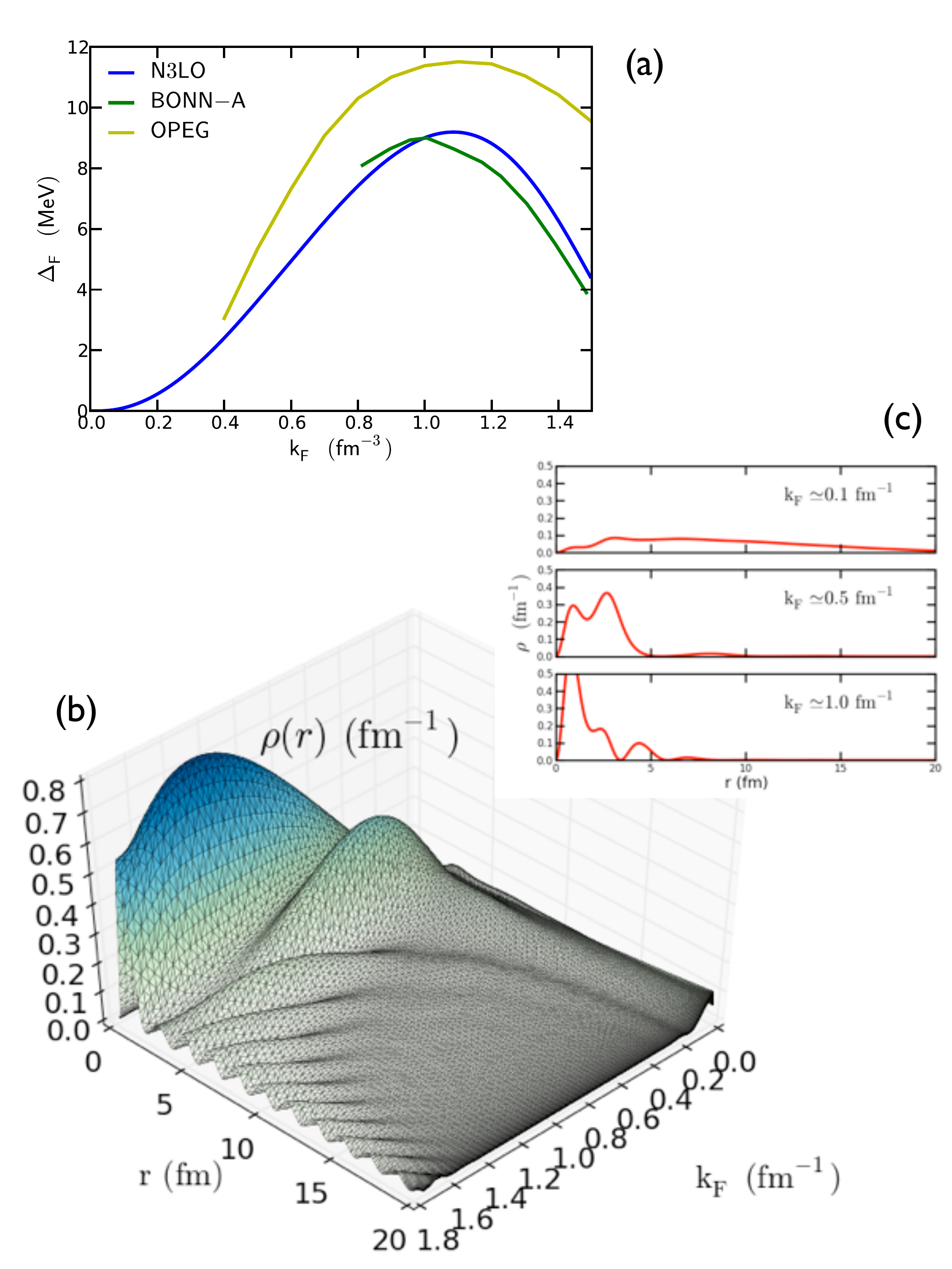}
\caption{Same as in Fig. \ref{fig2} for the $^3$SD$_1$ channel. In the upper panel {\bf (a)} we include as benchmarks some well known results (we refer the reader to Ref. \cite{Jensen} for more details).}\label{fig3}
\end{figure}

\begin{figure}[t]
\centering
\includegraphics[width=1.0\textwidth,natwidth=610,natheight=642]{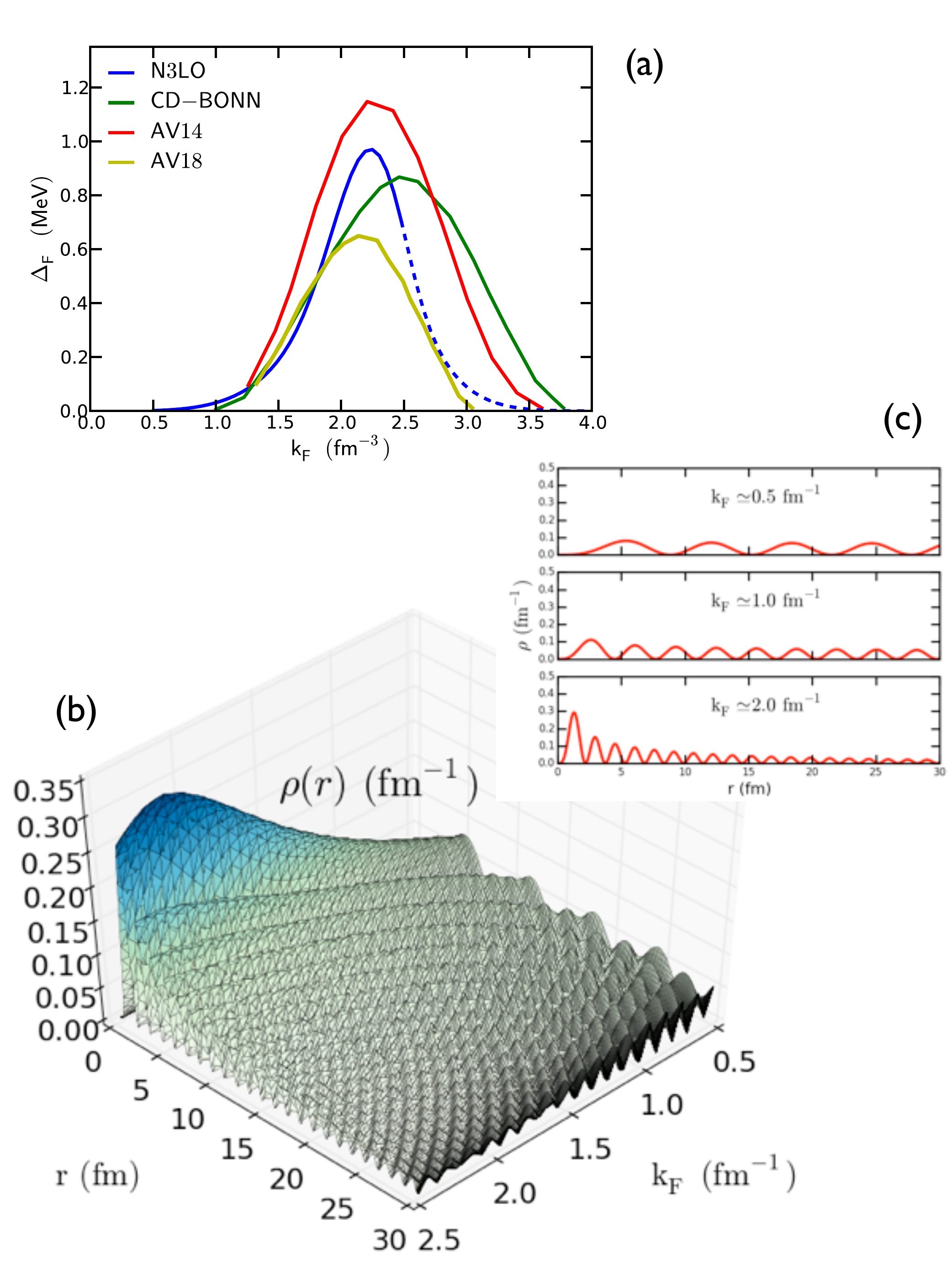}
\caption{Same as in Fig. \ref{fig2} for the $^3$PF$_2$ channel. In the upper panel {\bf (a)} we include as benchmarks some well known results (we refer the reader to Ref. \cite{Jensen} for more details).}\label{fig4}
\end{figure}

\paragraph{Triplet cases ($^3$SD$_1$ in nuclear matter and $^3$PF$_2$ in neutron matter)}
~\\
In Figs. \ref{fig3} and \ref{fig4}  we show the same informations as before, respectively for the $^3$SD$_1$ 
and $^3$PF$_2$ gaps. In the first case, our results seems to support the previous indication that the Cooper pair wave function merges smoothly into a "deuteron-like" wavefunction as density decreases \cite{bcsbec3sd1} and for higher densities the system belongs to a BCS-BEC crossover region. 
In the latter case, it is very clear that the possibility of a BCS-BEC crossover is ruled out because $\rho(r)$ behaves exactly as expected in a BCS weak coupling regime.

Of course, the previous results represent only a very preliminary analysis that need to be refined and improved, i.e. including three-body forces and self-energy effects \cite{futurepaper}. 

\section{Conclusions}

We have presented preliminary calculations of the Cooper pair wavefunctions  in infinite matter
employing realistic two-body nuclear forces derived within the framework of chiral 
effective field theory. 
The BCS gap equation is solved employing Khodel's method, which is found 
to be numerically stable with respect to variations of Gaussian integration points, momentum cutoffs and the
scale factor.  

\section{Acknowledgments}
Work supported in part by US DOE Grant No.\ DE-FG02-97ER-41014. 

%



\end{document}